# Germanium-Vacancy Color Center in Diamond as a Non-invasive Temperature Sensor

*Jing-Wei Fan‡†◊, Ivan Cojocaru‡†, Joe Becker†, Abdulrahman Alajlan†, Sean Blakley†, Mohammadreza Rezaee†, Anna Lyamkina⊥, Yuri N. Palyanov└⊫, Yuri M. Borzdov└⊫, Ya-Ping Yang◊, Aleksei Zheltikov†⌐∥, Phil Hemmer†, Alexey V Akimov†∥⌐*

† Texas A&M University, Department of Physics and Astronomy, 4242 TAMU, College Station, TX, USA

◊ MOE Key Laboratory of Advanced Micro-Structured Materials, School of Physics, Science and Engineering, Tongji University, Shanghai 200092, China

∥ PN Lebedev Institute RAS, Leninsky Prospect 53, 119991, Moscow, Russia

⌐ Russian Quantum Center, 100A, Novaya Street, Skolkovo, 143025, Moscow, Russia

└ Sobolev Institute of Geology and Mineralogy, Siberian Branch of the Russian Academy of Sciences, Koptyug Ave., 3, 630090, Novosibirsk, Russia

⊫ Novosibirsk State University, 1, Pirogova str, 630090, Novosibirsk, Russia

⊥ A.V. Rzhanov Institute of Semiconductor Physics, Siberian Branch of the Russian Academy of Sciences, Pr. Lavrent'eva 13, 630090, Novosibirsk, Russia

| Physics Department, International Laser Center, M.V. Lomonosov Moscow State University,

Moscow 119992, Russia

ABSTRACT: We present high-resolution, all-optical thermometry based on ensembles of GeV color center in diamond. Due to the unique properties of diamond, an all-optical approach using this method opens a way to produce non-invasive, back-action-free temperature measurements in a wide range of temperatures, from a few Kelvin to 1100 Kelvin.

Understanding the thermal properties of a living organism is a long-standing problem, since temperature is the most fundamental factor regulating all chemical reactions in vivo[1–4]. Development of high-resolution temperature measurement techniques made it possible to question temperature function at the single-cell level, or even within the cell[5–8]. Addressing the thermal properties of specific organelles opens the door to new possibilities for understanding intra-cell chemistry. Living cells actively react to environmental changes in temperature and are likely to change their internal temperature during such processes as division, gene expression, enzyme reaction, and metabolism[9,10]. This temperature change should be relatively small and transient, however, due to a cell's strong interactions with its environment. Therefore, detecting this temperature change is quite challenging.

A number of intra-cell temperature mapping techniques have been suggested in recent years. For example, fluorescent nanogel was used in combination with time-resolved photon counting, enabling temperature sensitivity of better than 0.5 degree inside the cell[6]. Local measurements with an ultrathin thermocouple were demonstrated to have a similar level of sensitivity[7]. Another interesting technique is based on quantum dots and the dependence of their photoluminescence spectra on temperature[11]. Probably the most precise measurement of temperature inside the cell was achieved using NV color center in diamond[5,8]. Here, by controllably heating a cell via laser illumination of a gold nanoparticle, researchers were able to achieve the temperature resolution well below 0.1 degree. In addition to providing superior sensitivity, this method has a number of advantages related to the intrinsic properties of diamond: it is chemically and physically inert, is not porous and has low toxicity, and such a sensor does not affect the cell's functionality to the extent possible. In addition, the surface passivation of diamond nanoparticles is well developed[12–16] enabling selective attachment of nanodiamonds to specific organelles of the cell, and therefore, in vivo measurement of their temperature response under various conditions.

While an NV-center-based sensor provides the best temperature sensitivity, it has a considerable limitation due to the fact that it requires the application of microwave radiation to the diamond. Since microwave radiation cannot be focused below a cell's size, the cell must be exposed to a large dose. At the same time, microwave radiation can produce heating or otherwise influence cell chemistry[17,18]. A possible alternative for the NV center may be a SiV center[19] or novel GeV center, which has optical spectra dominated by a zero-phonon line, thus offering a possibility to measure temperature through detection of their spectral response. This better-developed SiV color center also has a narrow spectral line, but compared to the SiV center, the GeV center has a nearly unitary quantum efficiency that minimizes possible heating effects due to the cell's interaction with optical radiation.

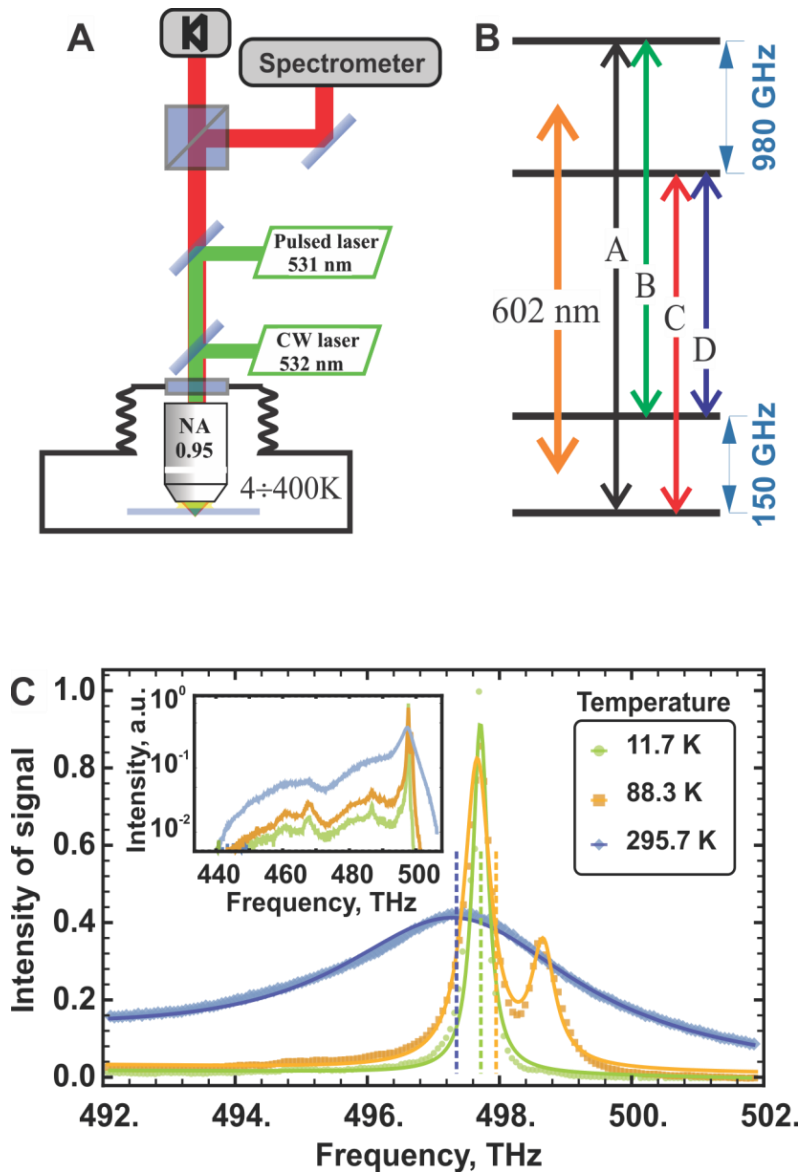

Figure 1. A) Schematic of the experimental setup. B) Level scheme of a GeV center. C) GeV center spectra at various temperatures. Solid lines correspond to fit, and dots, to experimental data. Dashed vertical lines represent the "center of mass" of the zero-phonon line.

The idea of GeV-based thermometry is founded on optical measurements of the shift of a spectral position of the zero-phonon line and its spectral width with the temperature change. The GeV center has a similar level structure to the SiV center[20–22]. Temperature dependence of SiV energy levels was thoroughly analyzed by Jahnke et al[23]. The shift and broadening of the spectral lines of an SiV center are dominated by the second- and first-order processes of the corresponding electron-photon interaction in the excited state. The rate of electron-phonon interaction is much higher than the spontaneous decay rate of the exited state in a wide range of temperatures, leading to the strong modification of the line width of the transition and the shift of its position. Since two-phonon processes dominate over single-phonon processes, this interaction results in $T^3$ dependence of the zero-phonon line position. To verify that the physics of the temperature shift of the GeV color center spectral line is similar, we measured a temperature dependence of the GeV zero phonon line.

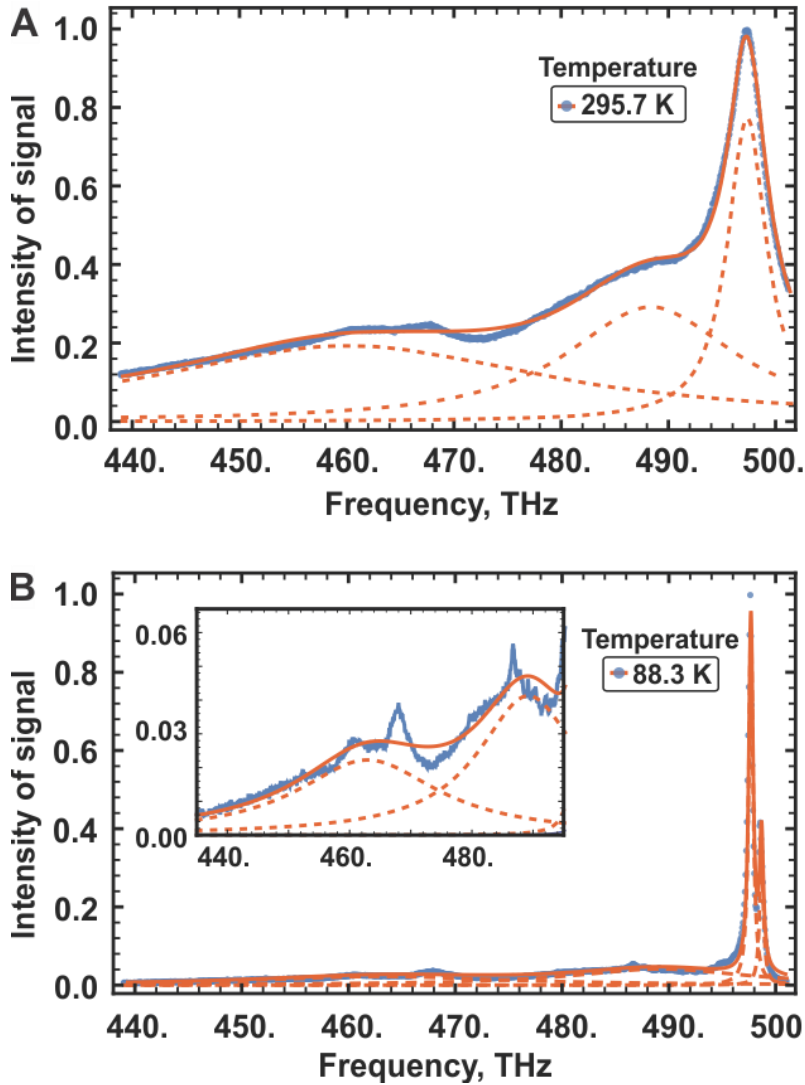

Figure 2. Fitting details. A) Fit with 3 Lorentzian curves at room temperature. Dashed lines indicate individual Lorentzian curves, and solid lines, overall fit. B) Fit with 4 Lorentzian curves at low temperature. Inset demonstrates sharp features at the spectrum, which appear at low temperatures. These features do not affect fits of the zero phonon line position and width.

For our measurements, we used a home-built, 2-channel confocal microscope (see Figure 1A). One of the channels was connected to an avalanche photodiode, allowing the sample visualization and focusing; the other was connected to the spectrometer (see Supporting Information for more details).

The zero-phonon line of the GeV center consists of 4 components, as shown in Figure 1B. Nevertheless, at room temperature, splitting between these 4 lines is completely covered by temperature-dependent broadening and, therefore, the band could be considered a single peak (see Figure 1C). Besides the zero-phonon line, the spectrum of the GeV center has a considerable fraction of emission to the phononic sideband. To take into account the presence of this sideband, we performed a fitting of the spectrum with 3 Lorentzian curves, as shown in Figure 2A.

With a temperature decrease, line broadening decreases, and about 150 K, the zero-phonon line splits into two peaks (Figure 3B & Figure 1C), corresponding to transitions A & B and C & D (Figure 1B). At this moment, fitting the spectrum with 3 Lorentzians becomes invalid. At temperatures lower than 150 K, 4 components/curves were used to account for the splitting of the zero-phonon line (see Figure 2B). To correlate the spectral position of the single line at higher temperatures with the doublet structure after splitting, one could calculate the "center of mass" of the doublet, as shown in Figure 1C. It is clear from Figure 3A that the "center of mass" of the doublet smoothly continues the dependence of the spectral shift of the single line on temperature, but deviates from the expected $T^3$ dependence. This is due to the fact that after the splitting, the relative intensities of the 2 lines in the doublet change with temperature (see Figure 1C). This change is rather easy to understand: the mixing of excited states by interaction with phonons leads to the rapid thermal relaxation of the exited stated population. Therefore, the ratio of spectral line intensities should be proportional to the Boltzmann factor $e^{-\Delta_e/kT}$ where $\Delta_e$ is the energy splitting in the excited state and $T$ is the temperature. Therefore, in order to calculate the "center of mass" of the doublet that takes temperature influence into account, one needs to include the Boltzmann factor as a weight of the spectral line (see Figure 3A). One could see that this correction is significant at temperatures below 150 K. Unfortunately, at temperatures right before the spectral line splits into doublet, correction of the weight using the Boltzmann factor is not possible. This is due to the reduced population of the upper level already affecting the observed position of the line, thus causing a slight deviation from the expected $T^3$ dependence right before the splitting temperature.

At the low temperature limit, the two peaks continue to shift apart (see Figure 3A) and change their relative amplitude, thus slowly converting to a single line due to the Boltzmann factor. In principle, one could expect further splitting of the remaining line into doublet corresponding to the ground state splitting, but due to the limitation of our spectrometer, this splitting was not observed. As noted, the theoretic model[23] predicts a cubic temperature dependence for the zero-phonon line width when temperatures are sufficiently high. Indeed, Figure 3B shows a dependence that is close to cubic, in agreement with the model used.

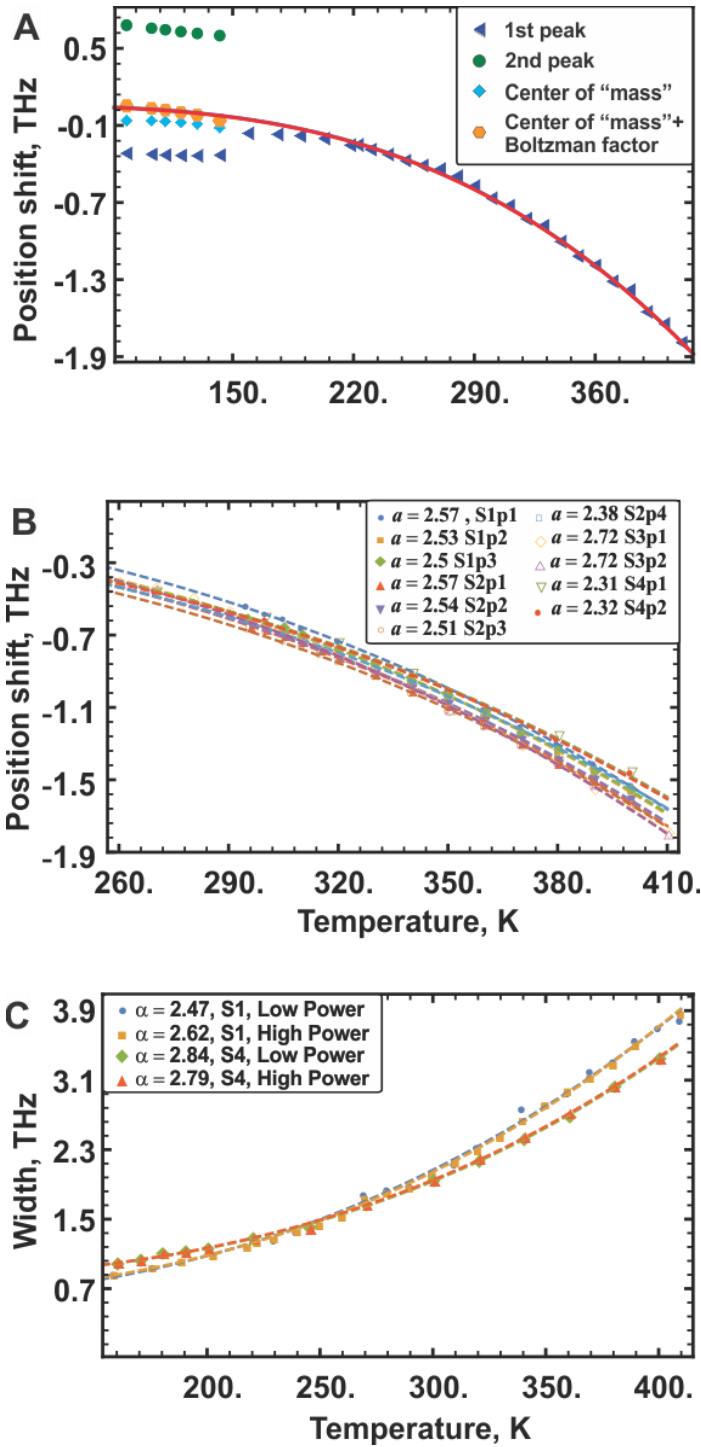

Figure 3. A) Temperature dependence of the zero-phonon line position for sample #1. B) Temperature dependence of the zero-phonon line position for various samples fitted with $aT^3 + b$ dependence. Index S stays for the sample, and p for position on the sample. C) Width of the zero

phonon line versus temperature fitted with $aT^\alpha + b$ dependence where $\alpha$ is used as a fit parameter. High power corresponds to 12 mw; low power corresponds to 1.4 mw.

To estimate an absolute uncertainty of our temperature measurements, we compared several samples with different concentrations of GeV centers and different methods of center creation. As one can see from Figure 3B, in all 4 cases, the temperature dependence is the same. Nevertheless, there is a small offset of the line position at room temperature that is most likely due to the different local strain, which leads to considerable uncertainty in the absolute temperature measurement. This uncertainty could be estimated as 0.5 K when one averages all the data taken.

To avoid this uncertainty, the single-point calibration technique could be used. In this technique, the position of the spectral line of GeV center is first measured for a known temperature, and then a relative change of temperature is measured by change in the position of the GeV line. Alternatively, one could use a GeV-based sensor only as a relative thermometer in a reasonable range of parameters. In the first case (see Figure 4A), uncertainty of the measured temperature could be estimated as 0.5 K per 10 seconds of integration time (see Supporting information), which is estimated as the standard deviation of measured points. The line width has a larger uncertainty (see Figure 4B) of 1 K. If calibrated by 1 point, the uncertainty of the absolute temperature measurement will be $0.5K + 0.1\delta T$ where $\delta T$ is detuning from the point where the sensor was calibrated. This uncertainty is mostly limited by the fitting accuracy at calibration temperature for shorter measurement times, and the spectrometer stability for longer times (see Supporting information) for temperatures around 300 K. At low temperatures, the main limitation factor is the resolution of our spectrometer. The upper limit of the range in which the sensor proposed is sensitive to the temperature is the ignition temperature of diamond, which is about 850º C.

On the other hand, the SiV case GeV center does not have any lifetime measurable temperature dependence. Our measurements, performed using a pulsed laser (see Figure 1A), are consistent with each other for all temperatures within 2% for samples 1 & 2 and within 6% for sample 3 & 4, which merely represents the uncertainty of lifetime measurements for these samples. The absence of such dependence clearly indicates that the decay of the exited state does not have a considerable photonic contribution and is purely radiative, which additionally minimizes the possible heating of an object of study by the thermometer based on GeV color centers.

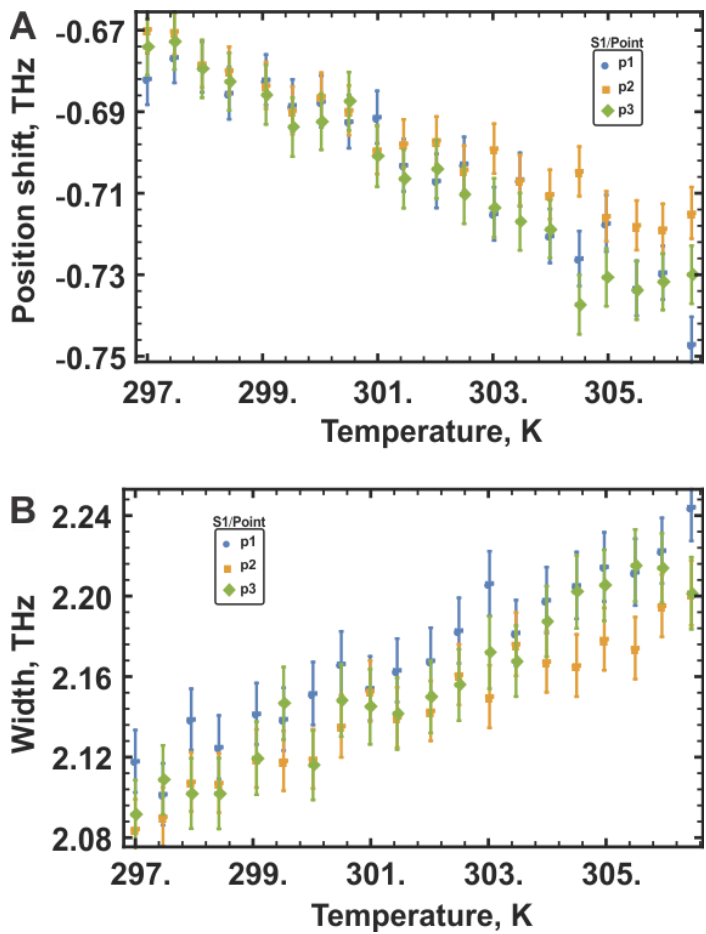

Figure 4. A) High-resolution temperature dependence of the GeV center zero-phonon line position. B) High-resolution temperature dependence of the GeV center zero-phonon line width.

We proved that the GeV spectral line depends on temperature via coupling of excited states to lattice phonons. Based on this mechanism/dependence, an all-optical, high-resolution thermometry technique that covers a wide temperature range from 4 K to 1100 K was proposed, and corresponding measurements from low to high temperature regions were demonstrated. In particular, this mechanism provides a non-invasive temperature sensor for biological applications, including intra-cell, in vivo measurements.

## ASSOCIATED CONTENT

**Supporting Information**.

The following files are available:

PDF, with details of experimental setup, samples used and fitting procedure

## AUTHOR INFORMATION

**Corresponding Author**

Alexey Akimov, akimov@physics.tamu.edu

**Author Contributions**

The manuscript was written through contributions of all authors. All authors have given approval to the final version of the manuscript. ‡These authors contributed equally.

## ACKNOWLEDGMENT

We thank Denis Sukachev, Alp Sipahigil, Christian Nguyen and Rufin Evans for fruitful discussions and samples #1 and #2.

This work was partly supported by the Russian Foundation for Basic Research (project nos. 16-02-00843, 17-52-53092), the Welch Foundation (Grant No. A-1801), the Russian Science Foundation (project no. 17-12-01533), ONR (00014-16-1-2578), National Natural Science Foundation of China (Grants No. 11474221, and No. 11574229), the Joint Fund of the National Natural Science Foundation of China (Grant No. U1330203), and International Exchange Program for Graduate Students, Tongji University. HPHT synthesis and characterization of Ge-doped diamond were accomplished with support from the Russian Science Foundation (Grant No. 14-27-00054).

ABBREVIATIONS

NV nitrogen-vacancy, SiV silicon-vacancy, GeV germanium-vacancy.

REFERENCES

(1)     Patel, D.; Franklin, K. A. *Plant Signal. Behav.* **2009**, *4* (7), 577–579.

(2)     Seymour, R. S. *Biosci. Rep.* **2001**, *21* (2), 223–236.

(3)     Warner, D. A.; Shine, R. *Nature* **2008**, *451* (7178), 566–568.

(4)     Bahat, A.; Tur-Kaspa, I.; Gakamsky, A.; Giojalas, L. C.; Breitbart, H.; Eisenbach, M. *Nat. Med.* **2003**, *9* (2), 149–150.

(5)     Kucsko, G.; Maurer, P. C.; Yao, N. Y.; Kubo, M.; Noh, H. J.; Lo, P. K.; Park, H.; Lukin, M. D. *Nature* **2013**, *500* (7460), 54–58.

(6)     Okabe, K.; Inada, N.; Gota, C.; Harada, Y.; Funatsu, T.; Uchiyama, S. *Nat. Commun.* **2012**, *3*, 705.

(7)     Wang, C.; Xu, R.; Tian, W.; Jiang, X.; Cui, Z.; Wang, M.; Sun, H.; Fang, K.; Gu, N. *Cell Res.* **2011**, *21* (10), 1517–1519.

(8) Fedotov, I. V; Safronov, N. A.; Ermakova, Y. G.; Matlashov, M. E.; Sidorov-Biryukov, D. A.; Fedotov, A. B.; Belousov, V. V; Zheltikov, A. M. *Sci. Rep.* **2015**, *5*, 15737.

(9) Clark, D. G.; Brinkman, M.; Neville, S. D. *Biochem. J.* **1986**, *235* (2).

(10) Lowell, B. B.; Spiegelman, B. M. *Nature, Publ. online 06 April 2000; | doi10.1038/35007527* **2000**, *404* (6778), 652.

(11) Wang, H.; Yang, A.; Sui, C. *Optoelectron. Lett.* **2013**, *9* (6), 421–424.

(12) Mochalin, V. N.; Shenderova, O.; Ho, D.; Gogotsi, Y. *Nat. Nanotechnol.* **2011**, *7*.

(13) Chang, B.-M.; Lin, H.-H.; Su, L.-J.; Lin, W.-D.; Lin, R.-J.; Tzeng, Y.-K.; Lee, R. T.; Lee, Y. C.; Yu, A. L.; Chang, H.-C. *Adv. Funct. Mater.* **2013**, *23* (46), 5737–5745.

(14) Hsu, T.-C.; Liu, K.-K.; Chang, H.-C.; Hwang, E.; Chao, J.-I. *Sci. Rep.* **2014**, *4*, S336–S344.

(15) Krüger, A.; Liang, Y.; Jarre, G.; Stegk, J. *J. Mater. Chem.* **2006**, *16* (24), 2322–2328.

(16) Say, J. M.; van Vreden, C.; Reilly, D. J.; Brown, L. J.; Rabeau, J. R.; King, N. J. C. *Biophys. Rev.* **2011**, *3* (4), 171–184.

(17) de la Hoz, A.; Díaz-Ortiz, Á.; Moreno, A. *Chem. Soc. Rev.* **2005**, *34* (2), 164–178.

(18) Kappe, C. O.; Stadler, A.; Dallinger, D. *Microwaves in organic and medicinal chemistry.*; Wiley-VCH, 2012.

(19) Nguyen, C. T.; Evans, R. E.; Sipahigil, A.; Bhaskar, M. K.; Sukachev, D. D.; Agafonov, V. N.; Davydov, V. A.; Kulikova, L. F.; Jelezko, F.; Lukin, M. D. *http://arxiv.org/abs/1708.05419* **2017**.

(20) Iwasaki, T.; Ishibashi, F.; Miyamoto, Y.; Doi, Y.; Kobayashi, S.; Miyazaki, T.; Tahara, K.; Jahnke, K. D.; Rogers, L. J.; Naydenov, B.; Jelezko, F.; Yamasaki, S.; Nagamachi, S.; Inubushi, T.; Mizuochi, N.; Hatano, M. *Sci. Rep.* **2015**, *5*, 12882.

(21) Palyanov, Y. N.; Kupriyanov, I. N.; Borzdov, Y. M.; Surovtsev, N. V. *Sci. Rep.* **2015**, *5* (October), 14789.

(22) Palyanov, Y. N.; Kupriyanov, I. N.; Borzdov, Y. M.; Khokhryakov, A. F.; Surovtsev, N. V. *Cryst. Growth Des.* **2016**, *16* (6), 3510–3518.

(23) Jahnke, K. D.; Sipahigil, A.; Binder, J. M.; Doherty, M. W.; Metsch, M.; Rogers, L. J.; Manson, N. B.; Lukin, M. D.; Jelezko, F. *New J. Phys.* **2015**, *17* (4), 43011.